\documentclass[dvips]{article}
\usepackage{supertabular,lscape,epsfig}
\usepackage{amssymb}
\usepackage{amsmath}
\DeclareSymbolFont{ppa}{OT1}{ppl}{m}{it}
\DeclareMathSymbol{\vv}{\mathalpha}{ppa}{'166}

\begin{document}

\newcommand{\dd}{\,{\rm d}}
\newcommand{\ie}{{\it i.e.},\,}
\newcommand{\etal}{{\it et al.\ }}
\newcommand{\eg}{{\it e.g.},\,}
\newcommand{\cf}{{\it cf.\ }}
\newcommand{\vs}{{\it vs.\ }}
\newcommand{\zdot}{\makebox[0pt][l]{.}}
\newcommand{\up}[1]{\ifmmode^{\rm #1}\else$^{\rm #1}$\fi}
\newcommand{\dn}[1]{\ifmmode_{\rm #1}\else$_{\rm #1}$\fi}
\newcommand{\upd}{\up{d}}
\newcommand{\uph}{\up{h}}
\newcommand{\upm}{\up{m}}
\newcommand{\ups}{\up{s}}
\newcommand{\arcd}{\ifmmode^{\circ}\else$^{\circ}$\fi}
\newcommand{\arcm}{\ifmmode{'}\else$'$\fi}
\newcommand{\arcs}{\ifmmode{''}\else$''$\fi}
\newcommand{\MS}{{\rm M}\ifmmode_{\odot}\else$_{\odot}$\fi}
\newcommand{\RS}{{\rm R}\ifmmode_{\odot}\else$_{\odot}$\fi}
\newcommand{\LS}{{\rm L}\ifmmode_{\odot}\else$_{\odot}$\fi}

\newcommand{\Abstract}[2]{{\footnotesize\begin{center}ABSTRACT\end{center}
\vspace{1mm}\par#1\par
\noindent
{~}{\it #2}}}

\newcommand{\TabCap}[2]{\begin{center}\parbox[t]{#1}{\begin{center}
  \small {\spaceskip 2pt plus 1pt minus 1pt T a b l e}
  \refstepcounter{table}\thetable \\[2mm]
  \footnotesize #2 \end{center}}\end{center}}

\newcommand{\TableSep}[2]{\begin{table}[p]\vspace{#1}
\TabCap{#2}\end{table}}

\newcommand{\FigCap}[1]{\footnotesize\par\noindent Fig.\  %
  \refstepcounter{figure}\thefigure. #1\par}

\newcommand{\TableFont}{\footnotesize}
\newcommand{\TableFontIt}{\ttit}
\newcommand{\SetTableFont}[1]{\renewcommand{\TableFont}{#1}}

\newcommand{\MakeTable}[4]{\begin{table}[htb]\TabCap{#2}{#3}
  \begin{center} \TableFont \begin{tabular}{#1} #4 
  \end{tabular}\end{center}\end{table}}

\newcommand{\MakeTableSep}[4]{\begin{table}[p]\TabCap{#2}{#3}
  \begin{center} \TableFont \begin{tabular}{#1} #4 
  \end{tabular}\end{center}\end{table}}

\newenvironment{references}%
{
\footnotesize \frenchspacing
\renewcommand{\thesection}{}
\renewcommand{\in}{{\rm in }}
\renewcommand{\AA}{Astron.\ Astrophys.}
\newcommand{\AAS}{Astron.~Astrophys.~Suppl.~Ser.}
\newcommand{\ApJ}{Astrophys.\ J.}
\newcommand{\ApJS}{Astrophys.\ J.~Suppl.~Ser.}
\newcommand{\ApJL}{Astrophys.\ J.~Letters}
\newcommand{\AJ}{Astron.\ J.}
\newcommand{\IBVS}{IBVS}
\newcommand{\PASP}{P.A.S.P.}
\newcommand{\Acta}{Acta Astron.}
\newcommand{\MNRAS}{MNRAS}
\renewcommand{\and}{{\rm and }}
\section{{\rm REFERENCES}}
\sloppy \hyphenpenalty10000
\begin{list}{}{\leftmargin1cm\listparindent-1cm
\itemindent\listparindent\parsep0pt\itemsep0pt}}%
{\end{list}\vspace{2mm}}

\def\TYLDA{~}
\newlength{\DW}
\settowidth{\DW}{0}
\newcommand{\dw}{\hspace{\DW}}

\newcommand{\refitem}[5]{\item[]{#1} #2%
\def\REFARG{#3}\ifx\REFARG\TYLDA\else, {\it#3}\fi
\def\REFARG{#4}\ifx\REFARG\TYLDA\else, {\bf#4}\fi
\def\REFARG{#5}\ifx\REFARG\TYLDA\else, {#5}\fi.}

\newcommand{\Section}[1]{\section{#1}}
\newcommand{\Subsection}[1]{\subsection{#1}}
\newcommand{\Acknow}[1]{\par\vspace{5mm}{\bf Acknowledgements.} #1}
\pagestyle{myheadings}

\newfont{\bb}{ptmbi8t at 12pt}
\newcommand{\xrule}{\rule{0pt}{2.5ex}}
\newcommand{\xxrule}{\rule[-1.8ex]{0pt}{4.5ex}}
\def\thefootnote{\fnsymbol{footnote}}
\begin{center}

{\Large\bf Mass Estimates for Some of the Binary Lenses in OGLE-III Database} 
\vskip 1.0cm
{\bf M.~~ J~a~r~o~s~z~y~\'n~s~k~i$^1$,~~ A.~~ U~d~a~l~s~k~i$^1$,~~ M.~~
K~u~b~i~a~k$^1$,\\
~~ M.\,K.~~ S~z~y~m~a~\'n~s~k~i$^1$,~~ G.~~ P~i~e~t~r~z~y~\'n~s~k~i$^{1.2}$,~~
I.~~ S~o~s~z~y~\'n~s~k~i$^{1,2}$,~~ \\
K.~~ \.Z~e~b~r~u~\'n$^1$,~~ O.~~ S~z~e~w~c~z~y~k$^1$~~
and~~ \L.~~ W~y~r~z~y~k~o~w~s~k~i$^1$}\\
\vskip 0.3cm
{$^1$Warsaw University Observatory, Al.~Ujazdowskie~4,~00-478~Warszawa, Poland\\
e-mail:\\
(mj,udalski,mk,msz,pietrzyn,soszynsk,zebrun,szewczyk,wyrzykow)@astrouw.edu.pl\\
$^2$ Universidad de Concepci{\'o}n, Departamento de Fisica, Casilla 160--C,
Concepci{\'o}n, Chile}
\end{center}

\Abstract{We model binary microlensing events OGLE 2003-BLG-170, 267, and 291.
Source angular sizes are measured for the events 267 and 291.  Model fits
to the light curves give parallaxes for the events 267 and 291, and
relative source sizes for 170 and 267. Selfconsistency arguments provide
extra limits on the models of the event 291. As a result we obtain
likelihood estimate of the lens mass for the event 170, mass measurement
based on angular size and parallax for 267, and narrow limits on mass in
the case of 291.  Brown dwarfs are most likely candidates for some of the
lens components.  The influence of the binary lens rotation and the Earth
parallax may be important but hard to distinguish when modeling relatively
short lasting binary lens events.}{Gravitational lensing -- Galaxy: bulge
-- binaries: general -- Stars: fundamental parameters}

\Section{Introduction}
Among several microlensing events discovered by the Early Warning System
(EWS - Udalski \etal 1994b, Udalski 2003a) of the third phase of the
Optical Gravitational Lens Experiment (OGLE-III) three binary lenses of
2003 season (2003-BLG-170, 2003-BLG-267 and 2003-BLG-291 -- compare
Jaroszy\'nski \etal 2004, hereafter Paper II) deserve a special
treatment. Events 170 and 267 have caustic crossings covered by
observations, which makes them possible candidates for mass
measurements. Event 291 is interesting due to its complexity: its light
curve might be understood as a result of a cusp approach followed by a
caustic crossing after several months.

The first microlensing phenomenon interpreted as being due to the binary
system was the event OGLE-7 (Udalski \etal 1994a). Several binary lens
events with good light curve coverage were used to study the atmospheres of
the source stars (\eg Albrow \etal 1999b) or to constrain the lensing
system parameters (\eg Alb\-row \etal 1999a). The first lens mass measurement
was obtained by An \etal (2002) based on a binary lens event with combined
effects of parallax motion and caustic crossing.

The lensing by two point masses was studied by Schneider and Weiss
(1986). Various aspects of binary lens modeling were described (among
others) by Gould and Loeb (1992), Bennett and Rhie (1996), Gaudi and Gould
(1997), Dominik (1999), Albrow \etal (1999c), and Graff and Gould (2002).
Some basic ideas for binary lens analysis can be found in the review
article by Paczy\'nski (1996).

In a typical situation one attempts to fit only the simplified
six\footnote{We count only the parameters defining the binary and the
source track. Another two parameters are necessary to give the source and
blend fluxes.} parameter binary lens models. Such models assume the source
to be point-like and neglect the effect of the Earth and binary orbital
motions. In fact majority of our models in the past (Jaroszy\'nski 2002 --
hereafter Paper I and Paper II) belonged to this class.

Due to the high data quality in cases of interest we are able to fit more
sophisticated models to the light curves, including the influence of the
source finite size, parallax effect, and, to some extend, the rotation of
the binary lens. These effects are not equally important in all three
cases, but we check all possibilities.

The calculation of lens magnification for extended sources, especially when
they cross caustics, is a time consuming numerical problem. The burden
becomes even heavier for events which have well covered caustic crossing
and so the most unpleasant case of calculations is required for high number
of source locations. We basically use the magnification calculation in the
lens plane (Dominik 1995, 1998, Bennett and Rhie 1996, Gould and Gaucherel
1997) following the numerical algorithms of Mao and Loeb (2001).

While using exact numerical methods is necessary in the final part of the
model optimization, in some cases one can use faster, approximate
calculation schemes to obtain solutions close to the optimal.

In the next Section we describe our models. Section~3 describes the three
events and shows the best fits to observations obtained under various
assumptions. The discussion follows in the last Section.

\Section{Technical Approach}
Our basic approach to modeling binary microlensing light curves is
described in Papers I and II. We repeat some of the definitions here for
completeness and because different groups tend to use somewhat different
parametrizations. Inclusion of parallax effects and binary rotation to our
models requires some refinements and introduction of extra parameters.

The search for solutions is based on  $\chi^2$ minimization  method for the 
light curves. It is convenient to model the flux at the time $t_i$ as: 
$$F_i =F(t_i)=A(t_i)\times F_s + F_b\eqno(1)$$
where $F_s$ is the flux of the source being lensed, $F_b$ the blended flux
(from the source close neighbors and possibly the lens), and the
combination $F_b+F_s=F_0$ is the total flux, measured long before or long
after the event. The lens magnification (amplification) of the source
$A(t_i)=A(t_i;p_j)$ depends on the set of model parameters $p_j$.  Using
this notation one has for the $\chi^2$:
$$\chi^2=\sum_{i=1}^N\frac{\left(A_i~F_s+F_b-F_i\right)^2}{\sigma_i^2}\eqno(2)$$
where $\sigma_i$ are the rescaled errors of the flux measurement taken from
the DIA photometry. The dependence of $\chi^2$ on the binary lens
parameters $p_j$ is complicated, while the dependence on the source/blend
fluxes is quadratic. The subset of equations $\partial \chi^2/\partial
F_s=0$; $\partial \chi^2/\partial F_b=0$ can be solved algebraically,
giving $F_s=F_s(p_j;\{F_i\})$ and $F_b=F_b(p_j;\{F_i\})$ thus effectively
reducing the dimension of the parameter space. In some cases this approach
may give unphysical solutions with negative blended flux ($F_b<0$). We
reject all such unphysical models as final results of minimization scheme;
paths to minima through unphysical regions of parameter space are not
prohibited.

The binary system consists of two masses $m_1$ and $m_2$, where by convention 
${m_1\le m_2}$. The Einstein ring radius of the binary lens is defined as: 
$$r_{\rm E}
=\sqrt{\frac{4G(m_1+m_2)}{c^2}~\frac{d_{\rm OL}d_{\rm LS}}{d_{\rm OS}}}
\equiv d_{\rm OL}\Theta_{\rm E}
\approx8~{\rm AU}~\sqrt{\frac{m_1+m_2}{M_\odot}~x(1-x)}\eqno(3)$$
where $G$ is the constant of gravity, $c$ is the speed of light, $d_{\rm
OL}$ is the observer--lens distance, $d_{\rm LS}$ is the lens--source
distance, and $d_{\rm OS}\equiv d_{\rm OL}+d_{\rm LS}$ is the distance
between the observer and the source, and $\Theta_{\rm E}$ is the angular
size of the Einstein ring. The approximate dependence of $r_{\rm E}$ on
binary mass and dimensionless distance to the lens $x\equiv
d_{\rm OL}/d_{\rm OS}$ is given for the future use. The distance
$d_{\rm OS}=8$~kpc is assumed in the calculation. The Einstein
radius serves as a length unit in the lens plane.

The binary system itself is described by two parameters: ${q{\equiv}
m_1/m_2}$ (${0{<}q{\le}1}$) -- the binary mass ratio, and $d$ -- binary
separation expressed in $r_{\rm E}$ units. If one neglects the binary
motion, its separation and orientation in the sky remain constant. The
influence of binary rotation is discussed below. We neglect the
possibility that the source itself belongs to a star system and assume that
it moves with constant velocity relative to the lens center of mass, as
observed by a heliocentric observer. Thus to a heliocentric observer the
source path projected into the lens plane is a straight line. One may
define: $b$ -- the impact parameter, the distance between the source
trajectory and the binary center of mass; $t_0$ -- the time of the source
passage by the center of mass; $\beta$ -- the angle between the source
trajectory and the line joining binary components measured at $t_0$;
$t_{\rm E}$ -- the Einstein time in which source travels a distance equal to
$r_{\rm E}$. The dimensionless source radius $r_{\rm s}\equiv
\Theta_*/\Theta_{\rm E}$
is the ratio of the source and the Einstein ring angular sizes.  This is
the complete list of seven binary lens parameters used in simplified
models.

The orbital motion of the Earth introduces an extra periodic component 
to the source motion relative to the lens. The amplitude of this effect
depends on the relative size of the  Earth orbit and the Einstein radius
projected into the observer's plane:
$$\pi_{\rm E} = \frac{1\mathrm{AU}}{{\tilde{r}}_{\rm E}}\quad{\rm where}
\quad{\tilde{r}}_{\rm E}\equiv 
r_{\rm E}~\frac{d_{\rm OS}}{d_{\rm LS}}\eqno(4)$$
The orientation of the source path as seen by a heliocentric observer
relative to the line of constant ecliptic latitude in the sky, $\psi$, has
an impact on the shape of the source trajectory. Sources in the Galactic
bulge have small ecliptic latitudes, so the parallax-induced motion along
the ecliptic strongly dominates its perpendicular component.

The binary lens orbit can be fully defined by six parameters giving its
size, eccentricity, orientation in space and rotation period. The fit of
all parameters would probably be possible only in cases, where the strong
source amplification lasts for time comparable to the revolution of the
binary. The binary projected into the sky is fully characterized by its
mass ratio, which remains constant, separation ($d$), and position angle
($\beta$). In the first order approximation one may assume the latter
parameters to change linearly with time:
$$\beta(t)=\beta_0+{\dot\beta}(t-t_0)\qquad d(t)=d_0+{\dot d}(t-t_0)\eqno(5)$$
where $d_0$, $\beta_0$ are the parameter values at the time
$t_0$, and  ${\dot d}$, ${\dot\beta}$ are their rates of change.

The simultaneous measurement of parallax effect and source angular radius
may be used to estimate the masses of the binary components (An \etal
2002). Using Eqs.~(3) and (4) and the definition of the dimensionless
source size $r_{\rm s}$ one has:
$$m_1+m_2=\frac{c^2}{4G}~{\tilde r}_{\rm E}~\Theta_{\rm E}
=\frac{c^2}{4G}~\frac{1~{\rm AU}}{\pi_{\rm E}}
~\frac{\Theta_*}{r_{\rm s}}\eqno(6)$$

Since the binary orbit parameters are not known, one can only roughly
estimate its orbital period. One has:
$$P_1 \approx 1~{\rm y}~\left(\frac{d~r_{\rm E}}{1~{\rm AU}}\right)^{1.5}~
          \left(\frac{m_1+m_2}{M_\odot}\right)^{-0.5}
P_2 \approx \frac{2\pi}{\dot\beta}\eqno(7)$$
respectively from the Kepler's IIIrd law or from the fitted rate of position
angle changes.

\Section{The Events -- Data and Modeling}
We use the OGLE III data, which are routinely reduced with difference
photometry based on algorithms developed by Alard and Lupton (1998) and
Alard (2000), giving high quality light curves of variable objects. The
Early Warning System (EWS) of OGLE III (Udalski 2003a) automatically picks
up candidate objects with microlensing-like variability. In some cases it
is possible to make more observations of objects of interest (several data
points per night), as compared to the usual survey mode (one per
night). The events we describe here belong to this class.  All our events
were already analyzed in Paper II using simplified (6 or 7 parameter)
models of binary lenses.

The large majority of observations were performed in $I$ filter only. The
field of 2003-BLG-291 has a few measurements in $V$ filter acquired in
2004, when the source was amplified. Other observations of fields of
interest in $V$ were obtained in 2005.

\subsection{OGLE 2003-BLG-170}
The event shows three distinct maxima in its light curve. The first may be
interpreted as a result of a cusp approach, and the remaining as a
following two caustic crossings with a characteristic "U-shaped" fragment
of the light curve between them. The caustic crossings are well
covered. The source is rather faint ($I_0=18.593$~mag) and its significant
(\ie stronger than 0.3~mag) lens magnification lasts for only 26 days. Due
to the proximity of a brighter star the measurement of the source angular
size based on color determination has not been possible in this case.

We have tried to improve our fit of the event light curve described in
Paper II using models including parallax and/or rotation. While some of the
models with parallax have formally lower $\chi^2$ compared to our best 7
parameter model, the difference is insignificant ($\le 0.5$). Apparently
neither parallax nor rotation could appreciably influence the light curve
of this short lasting event. Our calculations give slightly improved 7
parameter model of the event, very close to the one presented in Paper II.
\begin{figure}[htb]
\centerline{
\includegraphics[height=65mm,width=65mm]{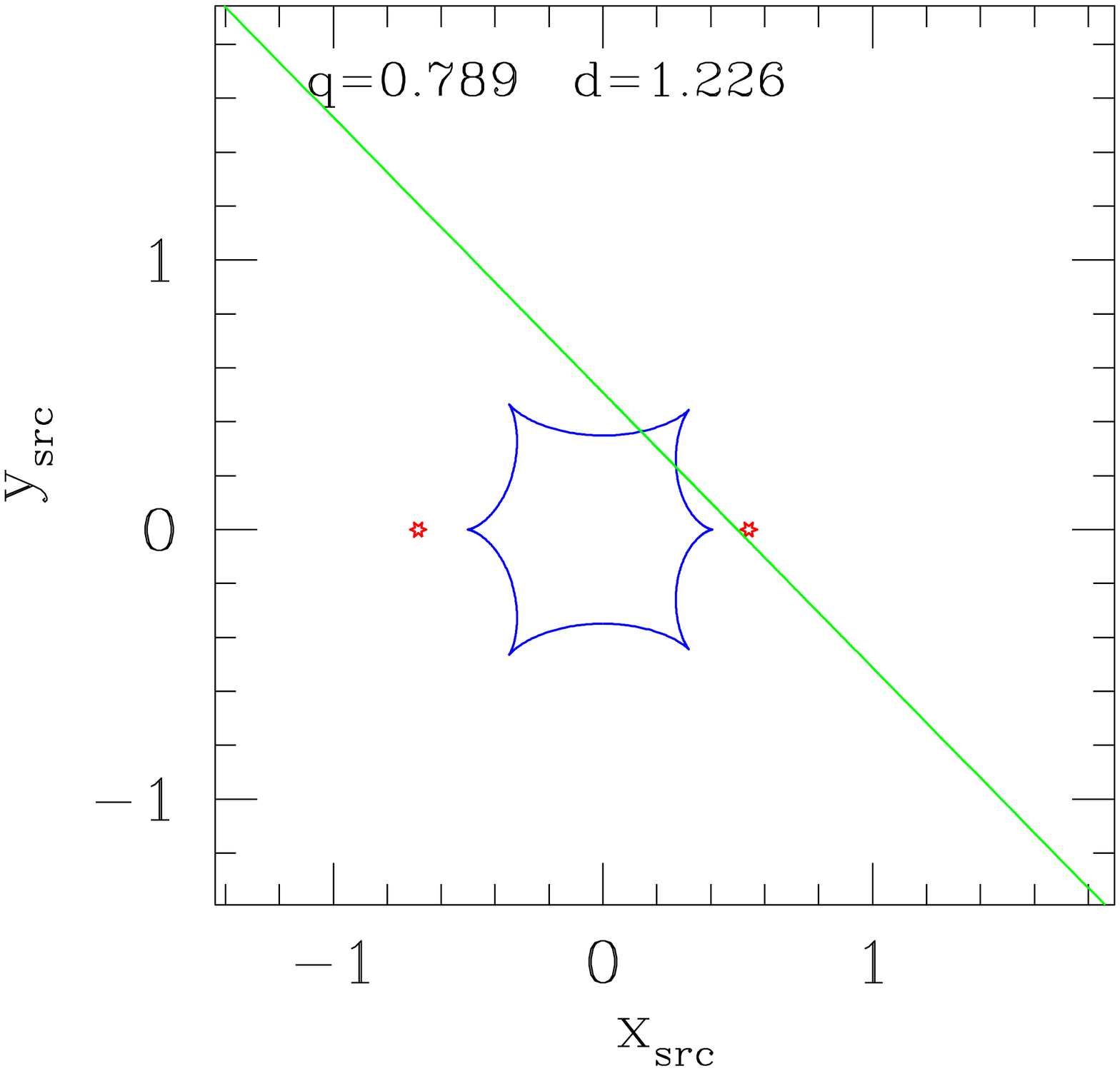} \hfill
\includegraphics[height=65mm,width=65mm]{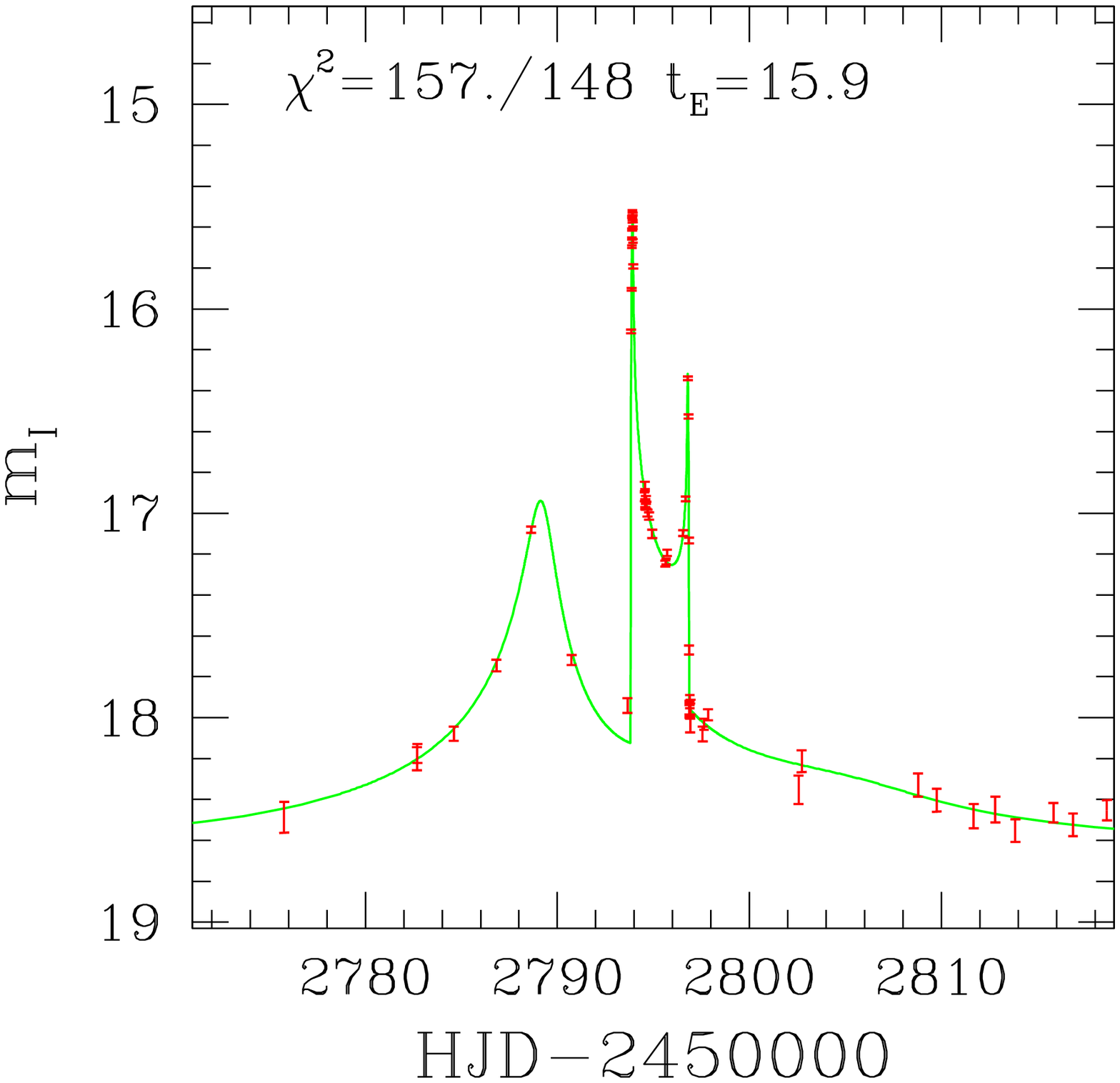}}
\vspace*{-3mm}
\FigCap{The standard (7 parameter) binary lens model of the event OGLE
2003-BLG-170.}
\end{figure}

The parallax measurement for this event is impossible, but the time
resolved caustic crossings observations allow to estimate the Einstein
radius projected into the source plane (Eq.~6). Our fit gives the flux
ratio $F_{\rm s}/F_0=0.735$, so the source apparent luminosity is $I_{\rm
s}=18.927\pm0.004$ and the blend has $I_{\rm b}=20.03\pm0.01$. Assuming the
source to be at the distance of 8~kpc and using OGLE-II extinction maps
(Udalski 2003b; Sumi 2004) we obtain $M_I=3.54\pm0.04$, where the
extinction is the main source of error. Thus the source star is $\approx
0.5$~mag brighter than the Sun. If it is a main sequence dwarf, it is
roughly 10\% larger than the Sun, and we estimate its radius to be
$R_{*}=(0.005\pm0.0005)$~AU. Using $r_{\rm s}=0.00267$ from the best fit
one gets the masses of the lens components as a function of the relative
lens distance $x\equiv d_{\rm OL}/d_{\rm OS}$. For the fitted mass ratio
$q=0.789$ one has:
$$m_1 + m_2 = (0.024~\MS+0.030~\MS)~\frac{x}{1-x}~\left(\frac{R_*}
{0.005~{\rm AU}}\right)^2~\frac{8~{\rm kpc}}{d_{\rm OS}}.\eqno(8)$$
The requirement that the lens should be fainter than $I_{\rm b}=20.03$ is
met for $x\le0.95$ ($d_{\rm OL}\le7.6$~kpc). 
The lens velocity perpendicular to the source--observer line,
$$\vv_\perp\equiv\frac{r_{\rm E}}{t_{\rm E}}=203~\frac{\rm km}{\rm s}
~x~\frac{R_*}{0.005~{\rm AU}}\eqno(9)$$
is typical for an object belonging to the Galactic disk. Using Han and
Gould (1996) model of the Galaxy we check the likelihood of measuring
\begin{figure}[htb]
\centerline{\includegraphics[height=85mm,width=85mm]{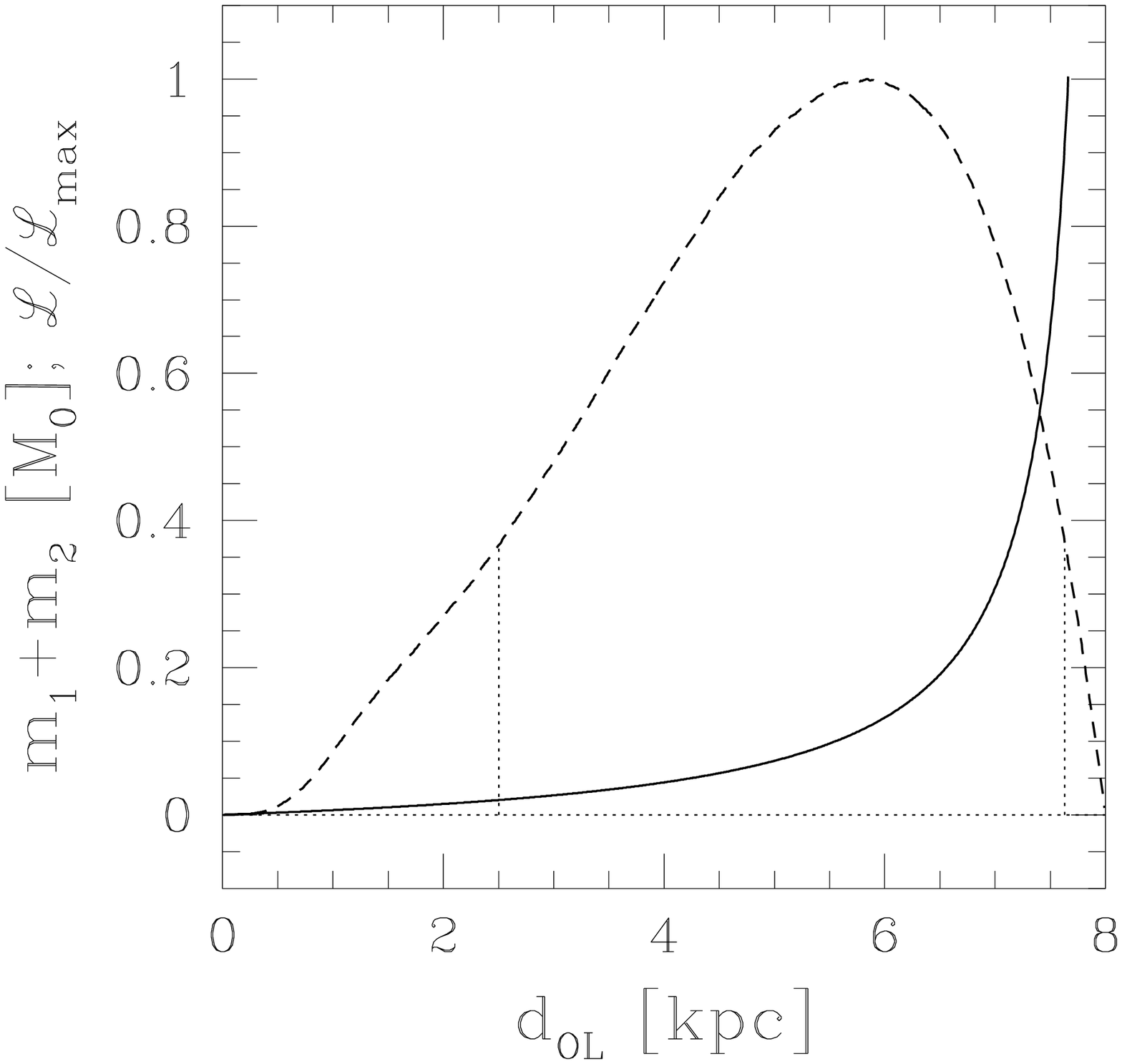}} 
\vspace*{-6mm}
\FigCap{The combined mass of the lens (solid line) and the likelihood 
function for its distance (dashed) for the event 2003-BLG-170. The mass
plot is continued to 7.6~kpc only, since larger lens distances are excluded
by the limit on lens luminosity. The source distance 8~kpc and size
0.005~AU are assumed. Dotted lines show the boundaries of the 90\%
likelihood region.}
\end{figure}
the obtained value of the relative lens velocity depending on the distance
to the lens. The observer's motion is a combination of Sun and Earth
motions. We take into account the peculiar velocity of the Sun and the
Earth velocity on June 6th (between caustic crossings). The lens and source
velocities are drawn at random from distributions adopted in the Galaxy
model for stars of the disk and bulge respectively.The relative likelihood
function for the source distance is shown in Fig.~2. The measurement of
relative velocity gives rather wide limits for the possible distance to the
lens: with 90\% probability it is $5.84^{+1.76}_{-3.34}$~kpc. The
corresponding masses of the lens components are:
$m_1=0.065^{+0.39}_{-0.054}~\MS$ $m_2=0.08^{+0.49}_{-0.061}~\MS$. Most
likely we deal with two brown dwarfs at the distance not exceeding
${\approx6}$~kpc, but dwarfs with masses ${\approx0.5~\MS}$ are not
excluded.

\subsection{OGLE 2003-BLG-267}
The light curve of 2003-BLG-267 has four distinct maxima, which were
observed between ${\rm HJD}=2452837$ and 2452852 (July/August 2003). This
shape may be interpreted (within binary lens hypothesis) as due to the
source cusp approach, two caustic crossings and another cusp
approach. The caustic crossings are well covered. The cusp approaches are
not particularly close, so the source remains 1 to 2~mag fainter compared
to its observed flux during the crossings. The amplification $\ge0.3$~mag
above the base brightness lasts for about 80 days. The angular radius of
the source based on ${V-I}$ \vs ${V-K}$ color relation of Bessell and Brett
(1988) and color--angular size calibration of van Belle (1999) could only
be obtained under the assumption that the blend is negligible. Fortunately
the result $\Theta_*=0.67\pm0.15~\mathrm{\mu as}$ is applicable to our
best models (see below) since the fits give practically vanishing blended
fluxes.

The standard 7 parameter modeling of the event gives a model light curve
with systematically displaced wings (\cf Fig.~3). This is true for both
close binary (model 1 in Table~1) and wide binary lenses (model 2). We do
not show any plots regarding wide binary models since they give much worse
fits of the light curve.
\begin{figure}[htb]
\centerline{
\includegraphics[height=64mm,width=64mm]{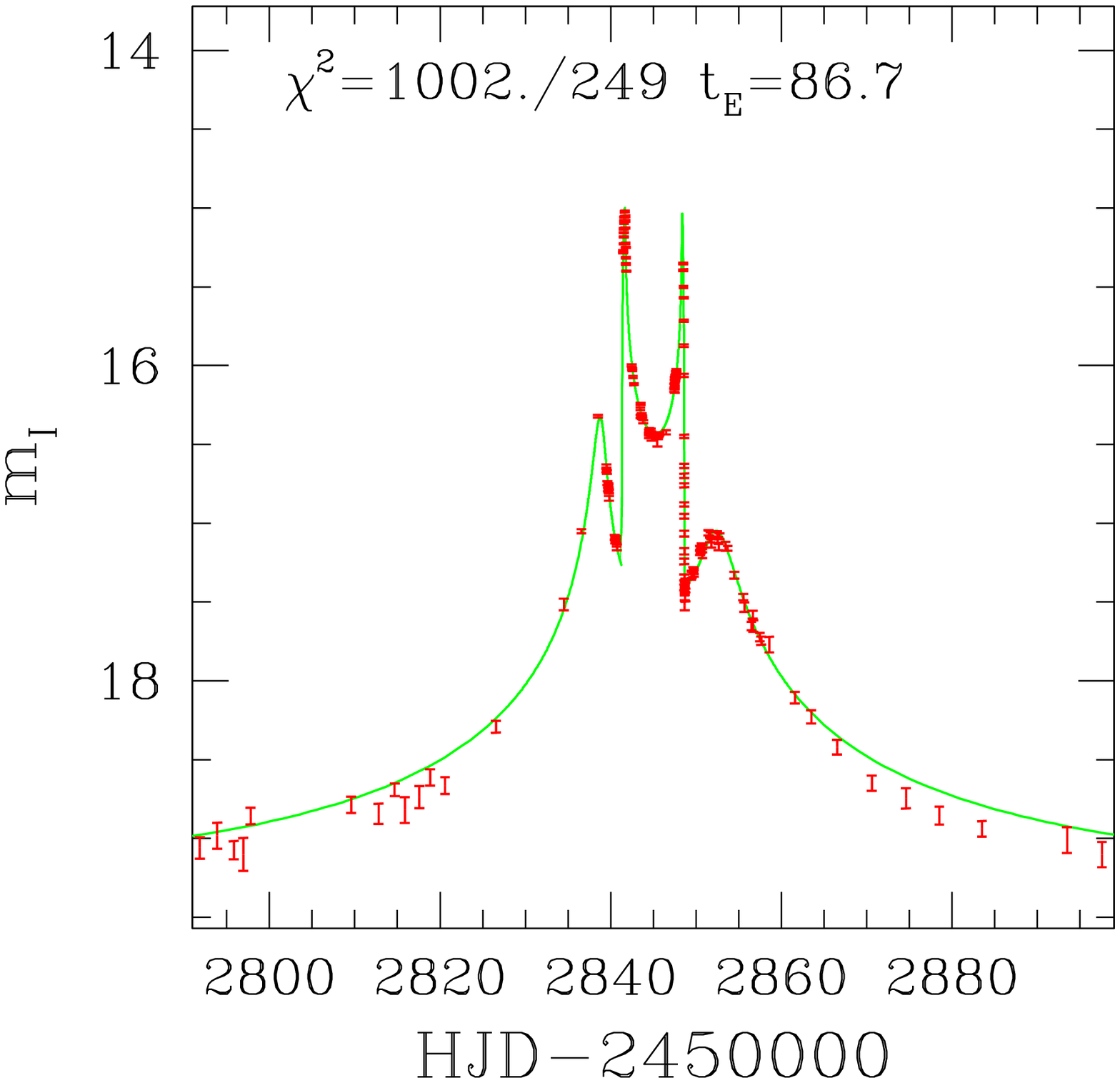} \hfill
\includegraphics[height=64mm,width=64mm]{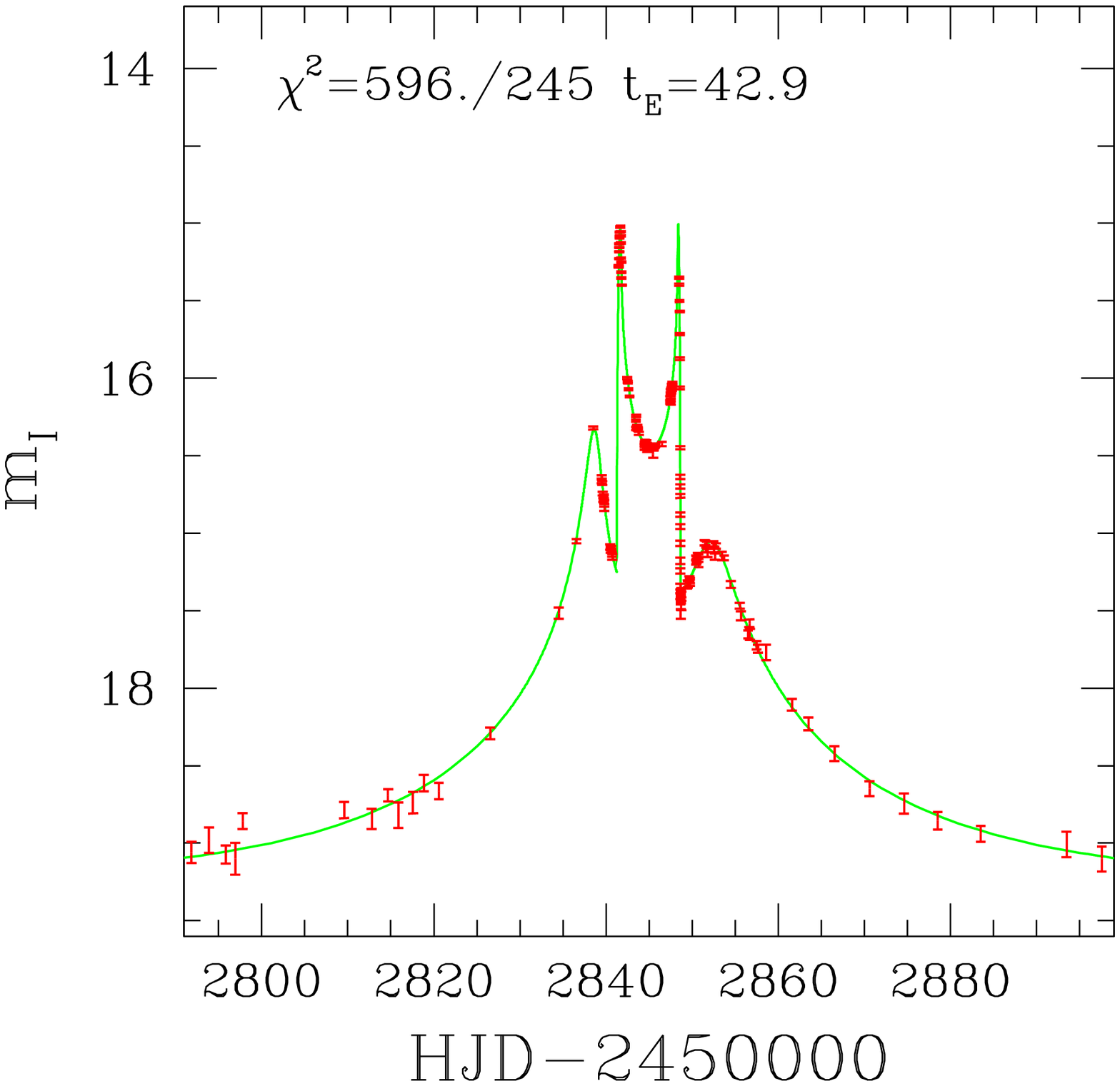}}
\vspace*{-5mm}
\FigCap{The comparison of fits to the light curve of the event OGLE
2003-BLG-267 using a standard, 7 parameter model of the binary lens ({\it
left}) and the improved, 7+4 parameter model taking into account parallax
and binary rotation.}
\end{figure}

The systematic differences between the model light curve and observations
suggest that some effects neglected in modeling are in fact
important. Therefore, we use models with parallax effect, binary rotation
and their combination what substantially improves the model quality, at
least for some close binary lenses. For the wide binary lenses the
inclusion of parallax and/or rotation helps in some cases, but the fits
remain worse than for the best 7 parameter model using a close binary.

The following figures show the source paths for various models. The
estimates of the binary component masses, the period of revolution and the
lens distance are also given in Table~1.

Models taking into account the Earth motion (but not the binary rotation)
improve the fit and give a high value of parallax. The best model of this
kind (model 3 in Table~1) predicts very small mass and rather short
rotation period of the binary. The estimated rotation rate of this model is
in fact higher than in the models taking rotation into account. Since the
lower rotation rate of the other models has a strong influence on the
modeling, our model is not selfconsistent, unless the binary is viewed at
very specific orientation and phase of the orbital motion -- a rather
unlikely situation.
\begin{figure}[htb]
\centerline{
\includegraphics[height=64mm,width=64mm]{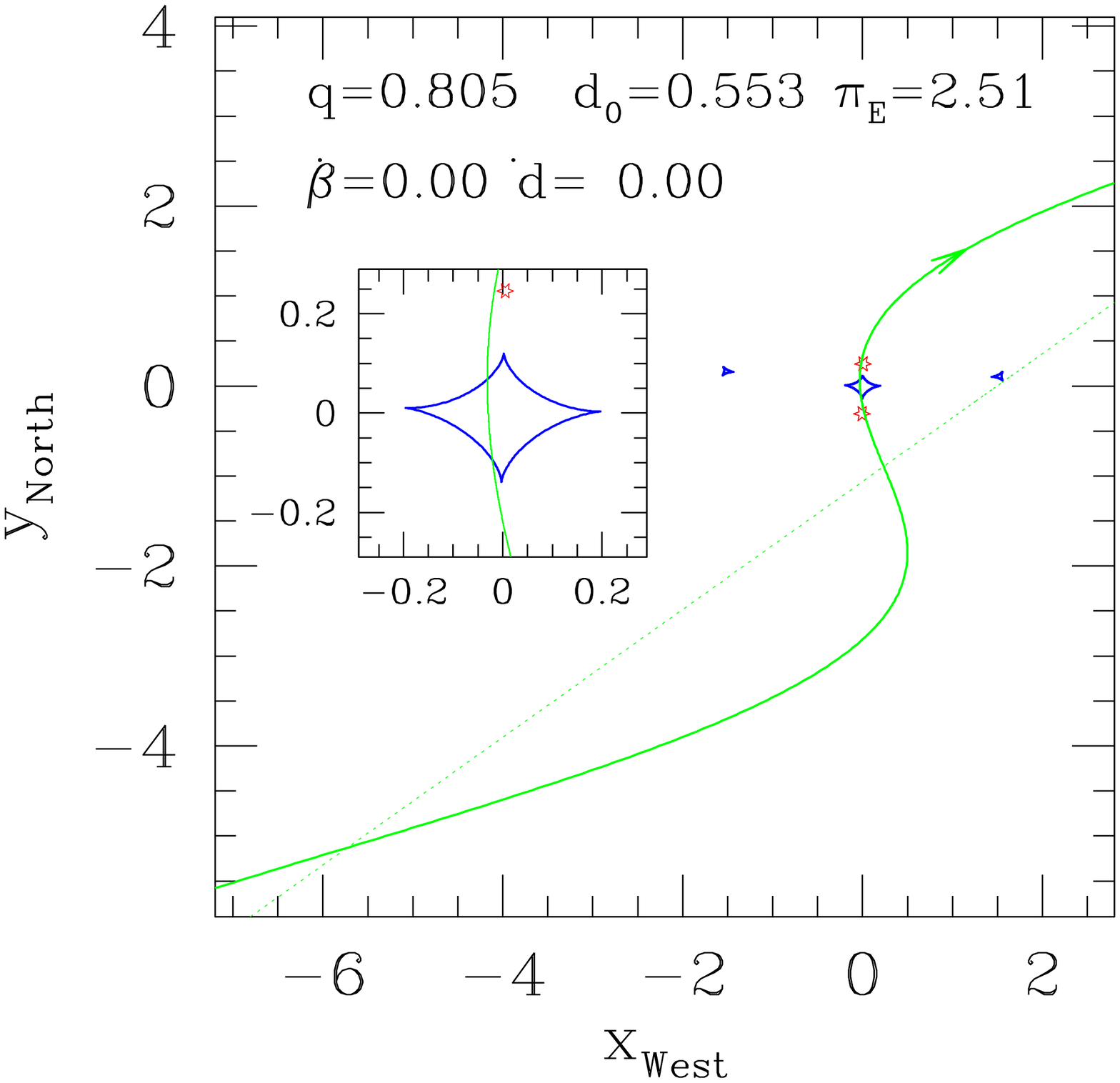} \hfill
\includegraphics[height=64mm,width=64mm]{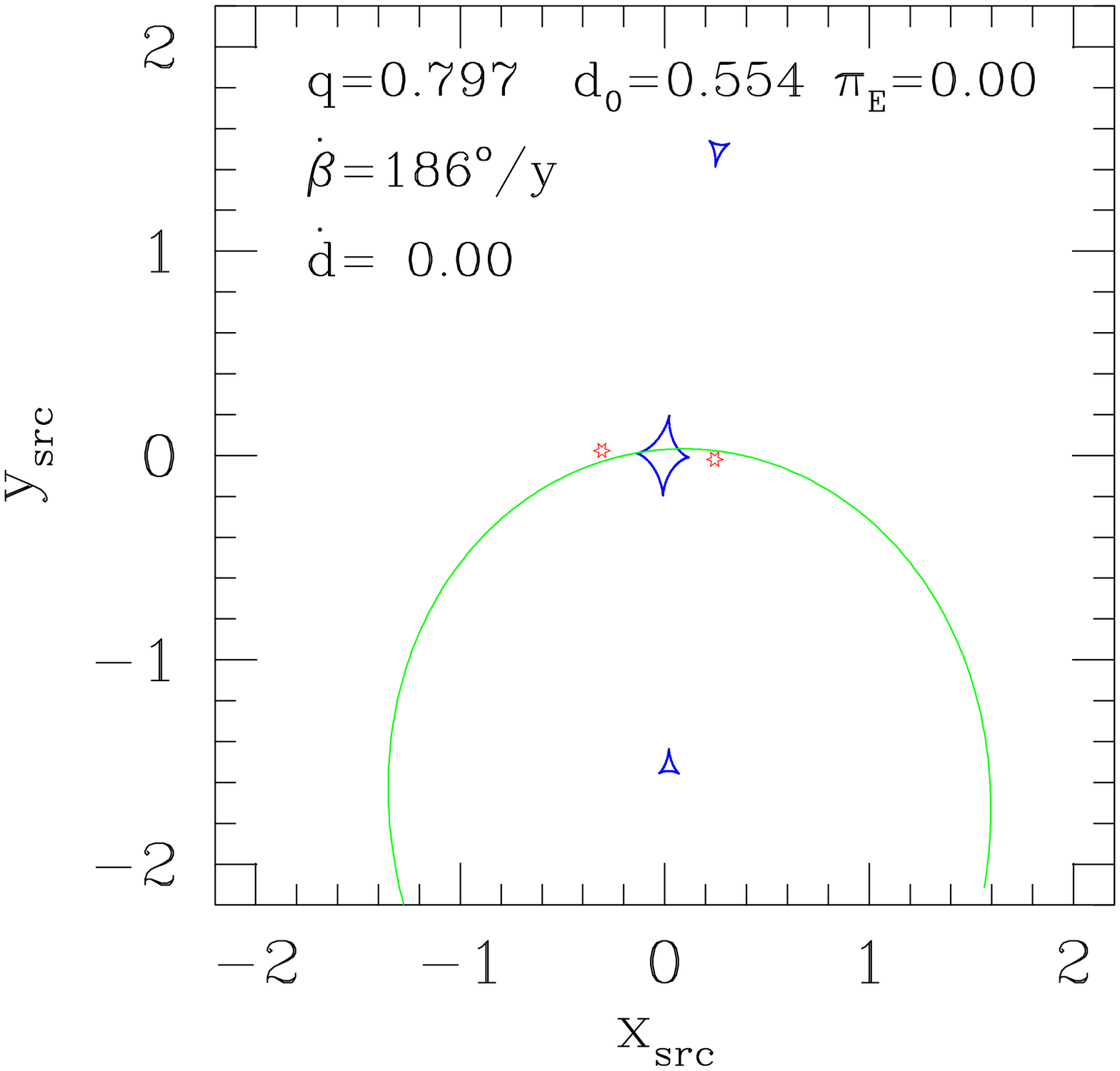}}
\vspace*{-3mm}
\FigCap{OGLE 2003-BLG-267: the source tracks for 7+2 parameter models   
including parallax effect ({\it left}) or rotation ({\it right}). On the
{\it left} the source trajectory is drawn as seen by an observer from
Earth, on the {\it right} -- in the frame of rotating binary.}
\end{figure}

The models with rotation (but no parallax) represent another improvement
compared to the standard approach. The direct lens mass estimate is not
possible in this case. The fit of the relative source size and measurement
of its angular size may be used to estimate the Einstein radius itself (and
the binary component masses) as a function of the lens distance. This
allows to derive another binary period estimate. The projection and
orientation effects make the simple comparison of the two rotation rate
estimates inconclusive.
\begin{figure}[htb]
\centerline{\includegraphics[height=95.0mm,width=95.0mm]{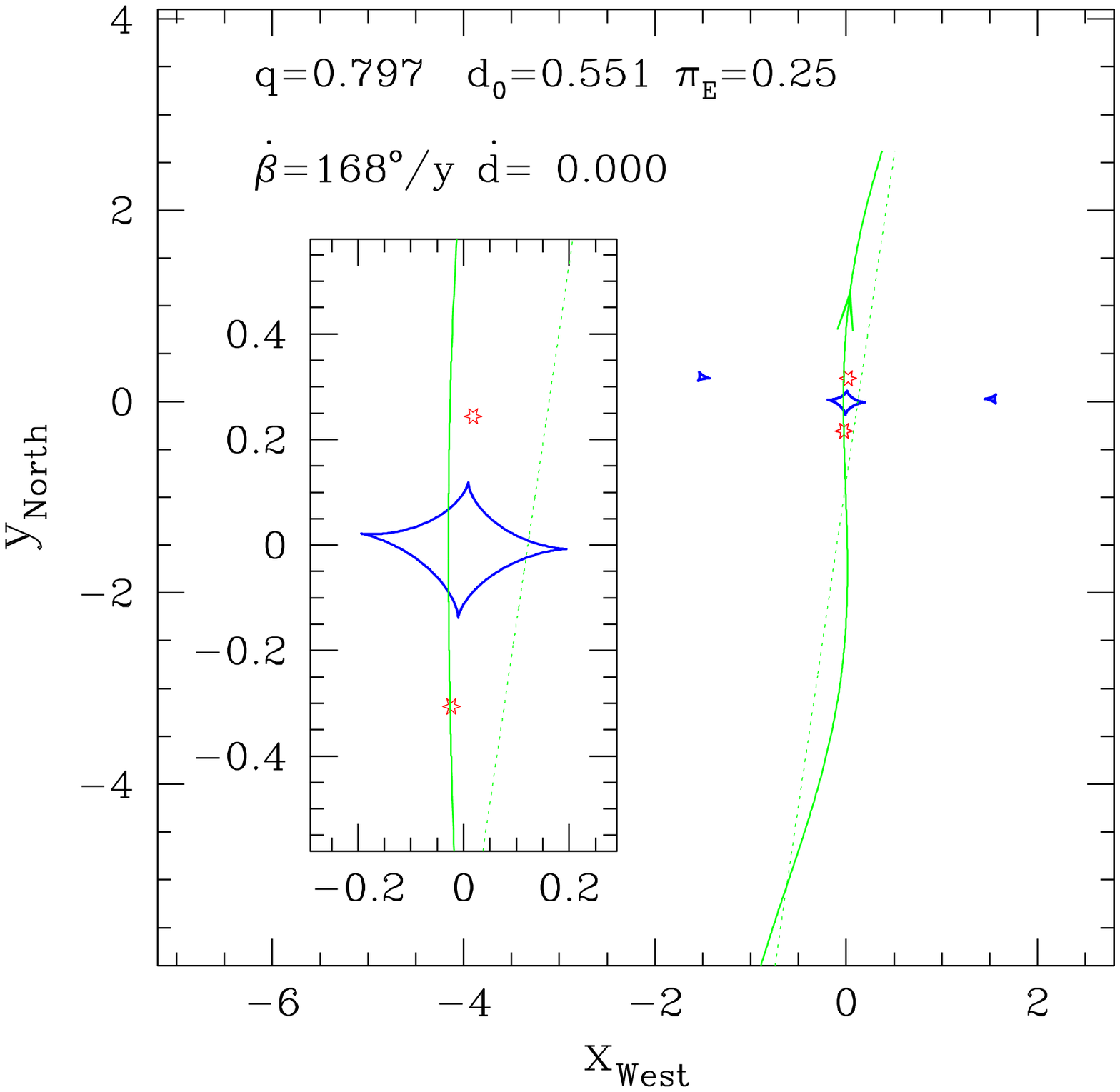}}
\vspace*{-7mm}
\FigCap{OGLE 2003-BLG-267: track of the source for the best model 
including rotation and parallax (7+4 parameters). (Solid line: as seen
from Earth, dotted line: as seen by a heliocentric observer).}
\end{figure}
\begin{figure}[htb]
\centerline{\includegraphics[height=95.0mm,width=95.0mm]{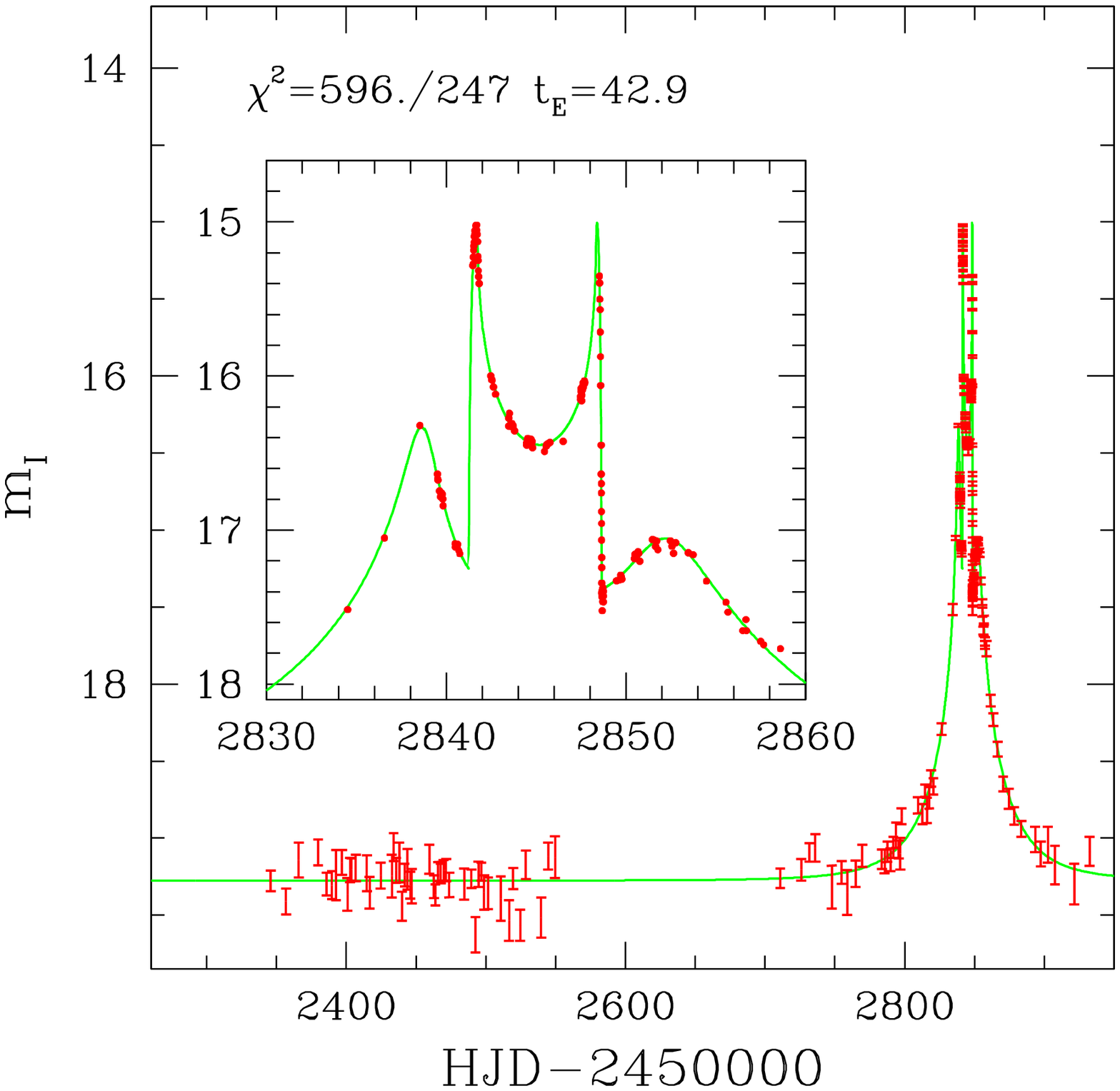}}
\vspace*{-8mm}
\FigCap{OGLE 2003-BLG-267: the light curve for the best model (7+4
parameters).}
\end{figure}

\MakeTable{c@{\hspace{3pt}}rcccccccccc}{12.5cm}
{Models of OGLE 2003-BLG-267}
{\hline
\noalign{\vskip3pt}
 & $\chi^2/$DOF & $q$ & $d_0$ & $r_s$ & $\pi_{\rm E}$ & ${\dot\beta}$ & $m_1$ & $m_2$ & $P_1$ & $P_2$ & $d_{\rm OL}$ \\
 & & & $[r_{\rm E}]$ & $[r_{\rm E}]$ & & $[^\circ/\mathrm{y}]$ & $[M_\odot]$ & $[M_\odot]$ & $[\mathrm{y}]$ & $[\mathrm{y}]$ & $[\mathrm{kpc}]$ \\
\noalign{\vskip3pt}
\hline
\noalign{\vskip3pt}
1 & 1002./249 & 0.668 & 0.353 & 0.00085 & - & - & - & - & - & - & -\\
2 & 1170./249 & 0.159 & 5.020 & 0.00038 & - & - & - & - & - & - & -\\
3 &  608./247 & 0.805 & 0.553 & 0.00267 & 2.51 & - & 0.0048 & 0.0060 & 0.73 & - & 1.49\\
4 &  604./247 & 0.796 & 0.554 & 0.00269 & - & 186 & - & - & - & 1.88 & -\\
5 &  596./245 & 0.797 & 0.551 & 0.00266 & 0.25 & 168 & 0.052 & 0.065 & 1.66 & 2.17 & 5.44 \\
\noalign{\vskip3pt}
\hline
\noalign{\vskip7pt}
\multicolumn{12}{p{12.5cm}}{Note: The table gives some of the model parameters
and the results of mass, period and distance estimates where applicable.
Models 1 (a close binary) and 2 (a wide binary) give the best fits obtained
under the approximation excluding the effects of parallax and binary
rotation. Model 3 allows for parallax, but not for rotation; model 4
includes rotation, but no parallax effect, and model 5 takes into account
both effects.}}

Models taking into account parallax and/or rotation give substantially
better fits compared to standard seven parameter models. There is an
elongated region in $(\pi_{\rm E},\dot\beta)$ plane running along the straight
line given by 
$$\frac{\pi_{\rm E}}{2.51}
+\frac{\dot\beta}{186\arcd/{\rm y}}\approx1\eqno(10)$$
which contains the high quality models. The best fit has parallax $\pi_{\rm
E}=0.25$ and rotation ${\dot\beta}=168\arcd~{\rm y}^{-1}$. Since there
are many local minima of $\chi^2$ in the parameter space, the confidence
regions for $\pi_{\rm E}$ are complicated. The 68\% confidence region is
limited to the close vicinity of the best fit. The 95\% confidence region
includes also a separate part with $\pi_{\rm E}\approx0$. Finally 99\%
confidence region consists of several other pieces around $\pi_{\rm E}
\approx0.074$, 0.15, 0.35, and 0.63. Thus the mathematical modeling
gives rather wide limits on possible value of parallax parameter.

The flux ratio for the best fit model $F_{\rm s}/F_0=0.994$ is close to
unity and the same is true for all other models of comparable quality.
Thus we can make use of the estimate of the source angular size.

Assuming that the lens consists of two brown dwarfs or two main sequence
stars, and using dependence of the mass on the distance to the lens
(relation analogous to Eq.~8 for event 170), we estimate the apparent
luminosity of the lens. The strong limit on the lens flux puts an
interesting limit on the maximum distance to the lens ($d_{\rm OL}
\le 0.92~d_{\rm OS}$). Since the source size, lens location and the
parallax are related (Eqs.~4--6), this limit translates also into $\pi_{\rm
E}\ge0.05$. It also excludes high mass lens close to the source, unless it
consists of white dwarfs, neutron stars and/or black holes.

Using the best fit model we obtain the following estimate of the lens
component masses:
$$m_1+m_2=(0.055~\MS+0.068~\MS)~\frac{0.25}{\pi_{\rm E}}
~\frac{\Theta_*}{0.67~{\rm \mu as}}\eqno(11)$$
Formal (one sigma) error of parallax fit is very low for the best model
($\pi_{\rm E}=0.24974\pm0.00003$), so formally the above estimate gives
masses to $\approx22\%$, error resulting from imprecise knowledge of the
source size.

Using wide (99\%) confidence limits on parallax ($0.074 \le \pi_{\rm E} \le
0.63$) and neglecting the errors resulting from the source angular size, one
obtains safe limits on the masses: $m_1=0.055^{+0.13}_{-0.032}~\MS$ and
$m_2=0.068^{+0.16}_{-0.041}~\MS$. The corresponding limits on the lens
distance are: $d_{\rm OL}=5.44^{+1.59}_{-1.78}$~kpc.

Following An \etal (2002) we check the consistency of the model comparing
the transverse potential and kinetic energies of the binary.  Using their
formalism and notation we obtain $|T_\perp/K_\perp|=3.04$ for the best
model, where the potential energy $T_\perp$ is calculated using the
projected distance between the lens components instead of the true 3D
quantity and similarly the kinetic energy $K_\perp$ neglects the unknown
motion along the line of sight. Checking the value of projected energies
ratio for other models within 99\% confidence limits we find that
$|T_\perp/K_\perp|=4.02$ for $\pi_{\rm E}=0.074$, $5.83$ for $\pi_{\rm
E}=0.63$, and it always exceeds $3.0$ within the confidence region. All
these numbers are typical for a binary of almost every possible
eccentricity and orientation (\eg Fig.~12 in An \etal 2002).

The rotation of the best model is also consistent with our assumptions: the
position angle of the binary changes by $\approx37\arcd$ during 80 days
of substantial lens amplification, and by $\approx3\arcd$ between caustic
crossings. Since the lens separation remains constant, our approximate
description of the binary remains valid.

\subsection{OGLE 2003-BLG-291}
Observations of 2003 season alone could be well modeled as a single mass
microlensing event. Including the data from the beginning of 2004 season,
one could think of a double source model for the light curve. Finally, the
rapid rise of the flux after ${\rm HJD}=2453093$ (2004 March 29) and
characteristic behavior during the following drop of the brightness,
resemble the caustic crossing event. The light curve is not complete, only
the flux decrease after the first caustic crossing is covered by observations.

The observations in $V$ filter obtained in 2004 and 2005 seasons allow the
angular size measurement. Using the color--angular size relation we obtain
$\Theta_*=0.97 \pm 0.10~{\rm \mu as}$.
\begin{figure}[htb]
{\center
\includegraphics[height=125mm,width=125mm]{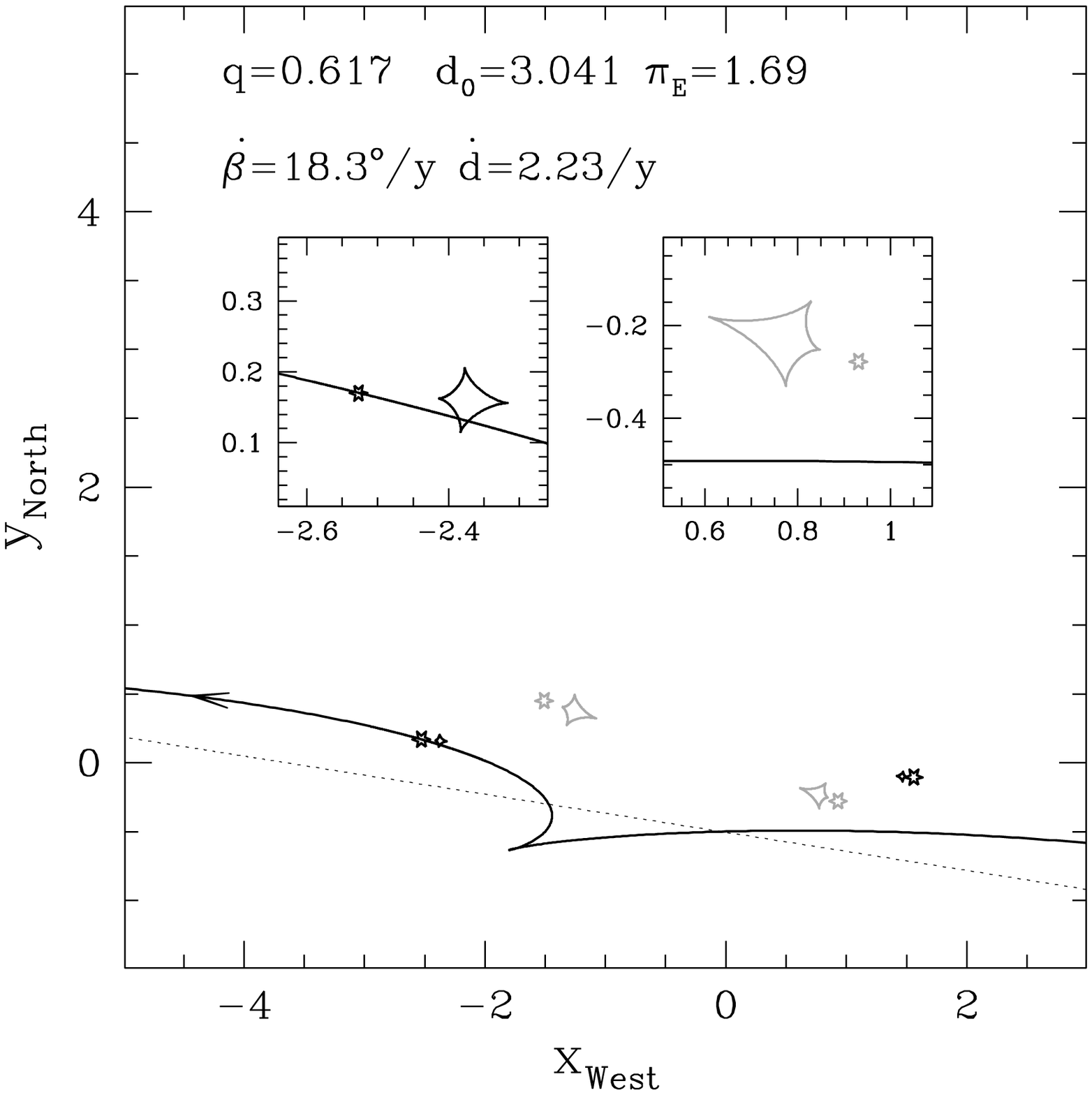}}
\vspace*{-12mm}
\FigCap{OGLE 2003-BLG-291: the model including parallax effect and binary
rotation (7+4 parameters). Source path projected into binary lens plane
is shown for seasons 2003 - 2004. (Dotted line: for the heliocentric
observer). The binary components positions and caustics are shown for 
the moment of maximum amplification in 2003 (grey) and during caustic
crossing in 2004 (black). Insert on the left shows the detail of
caustic crossings (2004) and insert on the right - the cusp approach (2003).}
\end{figure}

The maximum of brightness in 2003 season had taken place 255 days before
the caustic crossing in April 2004. Such a long time scale of the event
forces one to use models taking parallax into account. In fact the model
shown in Paper II was of this kind. We have not described it explicitly,
but the source trajectory shown there in Appendix 1 would never approach
the other caustic unless it were curved.
\begin{figure}[htb]
\centerline{
\includegraphics[height=127mm,width=127mm]{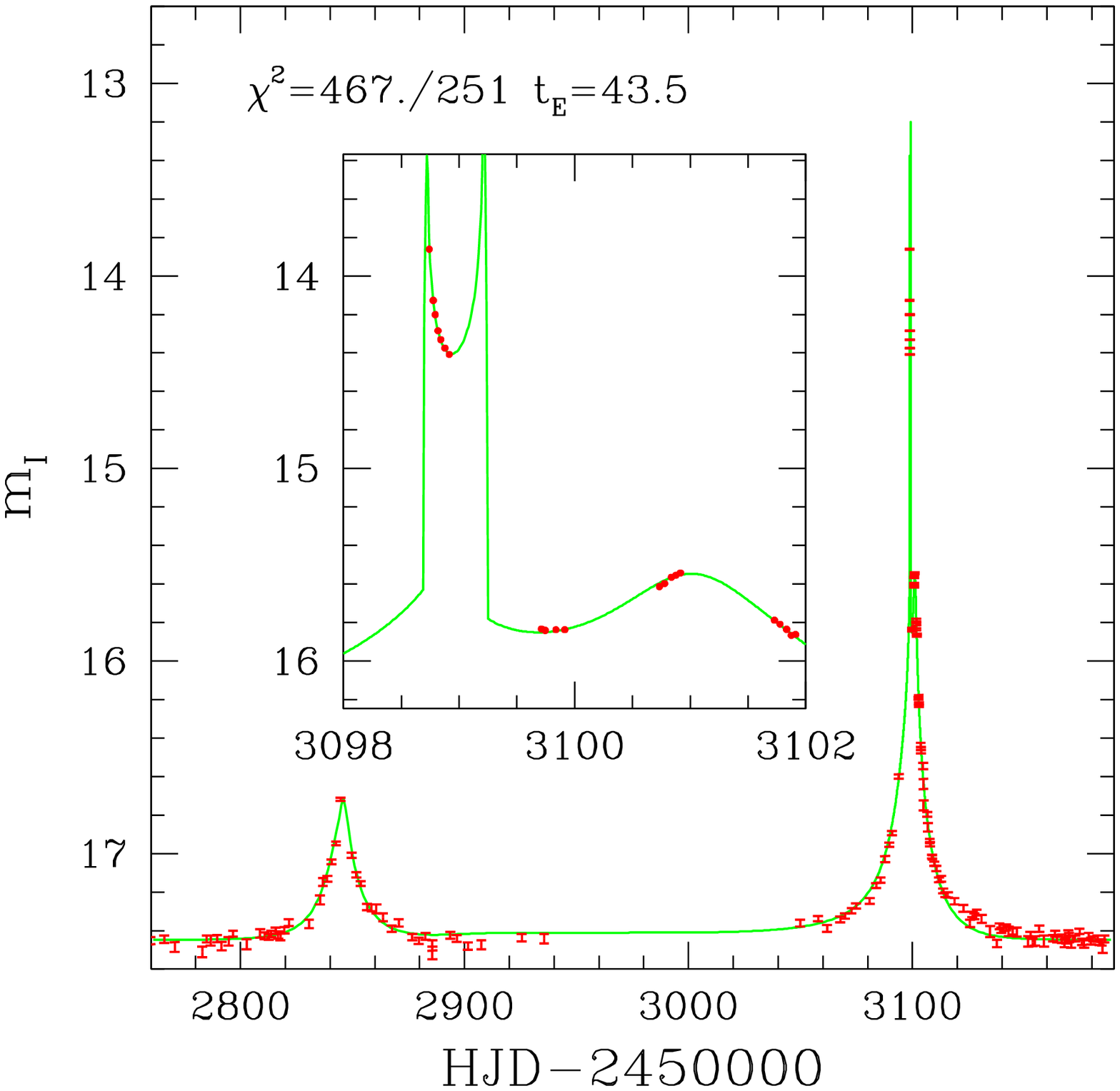}}
\vspace*{-11mm}
\FigCap{OGLE 2003-BLG-291: the light curve based on the model including 
parallax effect and binary rotation (solid line). Observations of parts of
2003 and 2004 seasons are shown (error bars). The insert shows the details
of caustic crossings.}
\end{figure}

When dealing with the event, we first try to model the observations from
the beginning of the 2004 season, with standard 7 parameter approach
tentatively using source trajectories crossing caustics. The best agreement
with observations is obtained for deltoid caustics of either close or wide
binary lenses and trajectories at an angle to the line joining lens
components. In the case of close binaries such lines may pass close to a
triangular caustic of the system for certain choices of mass ratios, but
the associated temporal amplification of the source is too short and/or too
weak to resemble the light curve of 2003.

Next we try 7+2 and 7+4 parameter approaches. We start from models with
deltoid caustics crossings and (using parallax and/or rotation) allow the
trajectories to bend. After quite extensive search over the parameter space
we obtain the best model shown in Figs.~7 and 8. This model includes
parallax with substantial influence of the binary rotation. The best model
with parallax but no rotation represents a significantly worse fit of the
event ($\chi^2$ larger by $>10^2$). We have not been able to obtain any
realistic model with rotation but no parallax; standard 7 parameter
approach is not capable of producing the event light curve which would even
qualitatively match the observations.

Both caustic crossings took place during the day, so a direct measurement
of the relative source size is not possible for 2003-BLG-291. The
minimization of $\chi^2$ chooses nonphysically small sources (practically
$r_{\rm s}\rightarrow0$). Small $r_{\rm s}$ implies however (\cf Eq.~6)
large angular size of the Einstein ring, small distance to the lens and its
high mass. Thus the requirement that the lens should not be too bright puts
a lower limit on the value of the parameter $r_{\rm s}$ for given values of
$q$, $\pi_{\rm E}$, $F_{\rm s}$, and $F_0$.

For the models including rotation we check the ratio of transverse
potential and kinetic energies (An \etal 2002) at the two epochs of
interest, when the source approaches or crosses the caustic. This criterion
makes models with relatively high velocities of binary components
physically uninteresting and we do not consider them here. We choose the
best of the physically plausible fits as our favored model. Its parameters
with standard errors are given in Table~2. The physical limitations may
introduce asymmetries to error estimates of some quantities, but we neglect
this effect.

\MakeTable{cccccccccccc}{12.5cm}{The best model of OGLE 2003-BLG-291}
{\hline
\noalign{\vskip3pt}
 $q$ & $d_0$            & $\beta_0$ 
   & $b$              & $t_0$ & $t_{\rm E}$ & $r_s$ 
     & $\pi_{\rm E}$ & $\psi$   
       & ${\dot d}$                  & ${\dot\beta}$        \\
     & $[r_{\rm E}]$ & [\arcd]  
   & $[r_{\rm E}]$ & [d]   &      [d]       & [$\times 10^{-3}$]       
     &                  & [$^\circ$]
       & [$r_{\rm E}/{\rm y}$] & [$\arcd/{\rm y}$] \\ 
\noalign{\vskip3pt}
\hline
\noalign{\vskip3pt}
0.617 & 3.04 & 184.7 & 0.50 & 2925.8 & 43.5 & 0.488 & 1.69 & 172.1 & 2.2  & 18.3\\
0.016 & 0.06 &   0.9 & 0.01 & 1.4    &  0.5 & 0.013 & 0.06 &   0.4 & 0.1  &  1.1\\
\noalign{\vskip3pt}
\hline
\noalign{\vskip3pt}
\multicolumn{12}{p{12.0cm}}{Note: The table gives all model parameters
and their estimated 1-$\sigma$ errors.}}
Using the best fit and the measurement of angular source size we get the
estimates of the binary lens components masses:
\setcounter{equation}{11}
\begin{eqnarray}
m_1 + m_2 & = & (0.056^{+0.002}_{-0.002}~\MS 
                +0.090^{+0.003}_{-0.003}~\MS)
                            ~\frac{\Theta_*}{0.97{\mu \rm as}} \\
          & = & 0.056^{+0.006}_{-0.006}~\MS
               +0.090^{+0.009}_{-0.009}~\MS
\end{eqnarray}
where the second equality takes into account also the error of the source
angular size measurement. The distance to the lens can be estimated as
$d_{\rm OL}=(0.29\pm0.03)$~kpc and the proper motion of the lens relative
to the source as $\mu_{\rm rel}=(16.7\pm 1.7)$~mas/y, errors resulting
mostly from the imprecise source size. The direction of the lens motion is
opposite to the source motion shown in Fig.~7. The linear velocity of the
lens, $\approx 23$~km/s is consistent with a peculiar velocity of an object
belonging to the Galaxy disk.

\Section{Discussion}
We have modeled three OGLE-III binary microlensing events of season 2003
trying to obtain estimates of the lens masses. Only in one case
(2003-BLG-267) the relative size of the source and parallax effect can be
fitted by modeling the light curve, and the angular size of the source is
measured using color--color relation, which gives masses of the binary
components. In the case of 2003-BLG-291 the parallax can be fitted and the
source angular size is measured. The relative source size cannot be fitted
directly from the optimization of the model, but the requirements that the
lens must not be too bright and the binary has to be a bound system, give
rather narrow limits on its value. The event 2003-BLG-170 allows only to
fit the source relative size. The measurement of the source angular
size can be to an extent replaced by the rough estimate of its physical
size, but it requires assumptions about the nature of the source and its
distance. Using this assumptions one can perform only a likelihood estimate
of the lens mass, using a model of velocity and density distributions in
the Galaxy.

In the case of 2003-BLG-267 we obtain binary lens components masses of
$0.055~\MS+0.068~\MS$. Within 95\% confidence limits the relative errors
are $\approx20\%$ and result mostly from imprecise knowledge of the source
size. Higher (99\%) confidence limits include substantial spread in
parallax estimate which becomes the main source of error. The limits on
masses become rather wide (0.023 to 0.19~\MS and 0.027 to 0.23~\MS,
respectively) including low mass main sequence stars. Still higher masses
are possible if the lens consists of compact faint objects like white
dwarfs, neutron stars and/or black holes. Such object would be located very
close to the source, which is extremely unlikely in the standard Galaxy
models.

In the case of 2003-BLG-291 we formally obtain the narrowest limits on the
lens component masses. One must however remember that our model is located
on the boundary of physically acceptable region in parameter space. This
means that the model is in a sense unnatural.  While our model binary lens
slowly changes its position angle in the sky, its separation changes are
rather rapid (see Fig.~7), which makes our approximate description of the
binary motion (${\dot d} \approx {\rm const}$) marginally
consistent. Probably the full modeling of the binary is necessary to
describe it selfconsistently, but such an approach is beyond the scope of
the present paper. It is also possible that the events of 2003 and 2004 are
caused by two separate lenses, one of them a binary. Since high
amplification of the source takes place at locations separated by few
Einstein radii, the third body responsible for the 2003 single event does
not have to influence strongly the binary responsible for the 2004 light
curve. (It is certainly model dependent, but large separation is at least
possible.) The most important argument against the three body hypothesis is
its low a priori probability: two independent microlensing events including
the same source have a chance $\approx p^2$, where $p\ll1$ is a chance of
one such event during a year. The triple lens is still another possibility
which has not been checked in our investigation.

Our lens mass estimate for OGLE 2003-BLG-267 is the third such attempt
(after An \etal 2002 and Kubas \etal 2005) using the parallax and source
size measurements for a binary microlensing event. In our approach we take
into account both parallax effect and source rotation in a way closely
resembling the method of An \etal (2002). The other group (Kubas \etal
2005) neglects rotation of the lens. In the case of 2003-BLG-267 both
effects play a role; to some extend one can be replaced with another which
is the main source of ambiguity of our parallax measurement. This is due to
the fact that the strong amplification lasts for rather a short time (80
days), so the periodic effects of parallax do not affect the light curve,
while the local bending of the trajectory can be accomplished using
rotation as well.

In some (but not all) binary microlensing events the parallax and binary
rotation may influence the important fragments of the source trajectory in
a way hardly distinguishable in the shape of light curves. We certainly
deal with such situation in our model of 2003-OGLE-267, where the quality
of fits using different mixtures of parallax and rotation is almost the
same. The effect is also present to some extent in modeling 2003-OGLE-291,
but in this case it involves other parameters as well, and some of the
concurrent solutions are physically invalid.  We deal with a degeneracy,
similar to better known degeneracies in single lens microlensing as
described by Smith, Mao and Paczy\'nski (2003) and Gould (2004). The binary
case is more complicated, since for a similar light curve one has to
reproduce a 2D trajectory of the source in the binary lens plane, instead
of 1D temporal dependence of its distance from the single mass. 

There are brown dwarfs candidates among binary lens components preferred by
our models. In all cases the higher mass alternatives are not excluded, but
in general our estimates give smaller lens masses compared to An \etal
(2002) -- $\approx0.6~\MS$ and Kubas \etal (2005) -- $\approx 0.5~\MS$. The
events caused by low mass lenses have typically shorter time scales and the
parallax effect should be rather difficult to discover, especially the
periodic seasonal modulation of the signal seen in some long lasting single
mass events (Mao \etal 2002, Smith \etal 2002). On the other hand the
parameter $\pi_{\rm E}$ is large for small Einstein radii, so the light
curves of low mass binary events have a greater chance to be significantly
influenced by parallax. It is not clear which of the contradicting effects
is more important, but taking into account the parallax in cases of even
short lasting binary events is worth a try.

\Acknow{We thank Bohdan Paczy\'nski for many helpful discussions 
and Shude Mao for the permission of using his binary lens modeling
software. This work was supported in part by the Polish KBN grants
2-P03D-016-24 and 2-P03D-021-24, the NSF grant AST-0204908, and NASA grant
NAG5-12212.}


\begin{references}
\refitem{Alard, C.}{2000}{\AAS}{144}{363}
\refitem{Alard, C., and Lupton, R.H.}{1998}{\ApJ}{503}{325}
\refitem{Albrow, M.D., \etal}{1999a}{\ApJ}{512}{672}
\refitem{Albrow, M.D., \etal}{1999b}{\ApJ}{522}{1011}
\refitem{Albrow, M.D., \etal}{1999c}{\ApJ}{522}{1022}
\refitem{An, J.H., \etal}{2002}{\ApJ}{572}{521}
\refitem{Bennett, D., and Rhie, H.}{1996}{\ApJ}{472}{660}
\refitem{Bessell, M.S. and Brett}{1988}{\PASP}{100}{1134}
\refitem{Dominik, M.}{1995}{\AAS}{109}{507}
\refitem{Dominik, M.}{1998}{\AA}{333}{L79}
\refitem{Dominik, M.}{1999}{\AA}{341}{943}
\refitem{Gaudi, B.S., and Gould, A.}{1997}{\ApJ}{486}{85}
\refitem{Gould, A.}{2004}{\ApJ}{606}{319}
\refitem{Gould, A., and Gaucherel, C.}{1997}{\ApJ}{477}{580}
\refitem{Gould, A., and Loeb, A.}{1992}{\ApJ}{396}{104}
\refitem{Graff, D.S., and Gould, A.}{2002}{\ApJ}{580}{253}
\refitem{Han, Ch., and Gould, A.}{1996}{\ApJ}{467}{540}
\refitem{Jaroszy\'nski, M.}{2002}{\Acta}{52}{39 (Paper I)}
\refitem{Jaroszy\'nski, M., Udalski, A., Kubiak, M., Szyma\'nski, M.,
Pietrzy\'nski, G. Soszy\'nski, I. \.Zebru\'n, K., Szewczyk, O.,
and Wyrzykowski, \L.}{2004}{\Acta}{54}{103 (Paper II)}
\refitem{Kubas, D., \etal}{2005}{\AA}{435}{941}
\refitem{Mao, S., and Loeb, A.}{2001}{\ApJL}{547}{L97}
\refitem{Mao, S., Smith, M.C., Wo\'zniak, P., Udalski, A., Szyma\'nski,
M., Kubiak, M., Pietrzy\'nski, G., Soszy\'nski, I., and \.Zebru\'n, K.}
{2002}{\MNRAS}{329}{349}
\refitem{Paczy\'nski, B.}{1996}{Ann. Rev. Astron. Astrophys.}{34}{419}
\refitem{Schneider, P., and Weiss, A.}{1986}{\AA}{164}{237}
\refitem{Smith, M.C., Mao, S., Wo\'zniak, P., Udalski, A., Szyma\'nski,
M., Kubiak, M., Pietrzy\'nski, G., Soszy\'nski, I., and \.Zebru\'n, K.}
{2002}{\MNRAS}{336}{670}
\refitem{Smith, M.C., Mao, S., and Paczy\'nski, B.}{2003}{\MNRAS}{339}{925}
\refitem{Sumi, T.}{2004}{\MNRAS}{349}{193}
\refitem{Udalski, A.}{2003a}{\Acta}{53}{291}
\refitem{Udalski, A.}{2003b}{\ApJ}{590}{284}
\refitem{Udalski, A., \etal}{1994a}{\ApJL}{436}{L103}
\refitem{Udalski, A., Szyma\'nski, M., Kaluzny, J., Kubiak, M., Mateo, M., 
Krzemi\'nski, W., and Paczy\'nski, B.}{1994b}{\Acta}{44}{227} 
\refitem{Van Belle, G.T.}{1999}{PASP}{111}{1515}
\end{references}
\end{document}